# Architectural Evolution and Selection Framework for Database Systems in AI-Ready Data Platforms

**Mohit Srivastava, Independent Researcher**

**Email:** mosriva@gmail.com. **ORCID ID:** 0009-0006-8692-1813.

**ABSTRACT** The rise of polyglot data management and AI-ready database architectures has created a complex design space across diverse database paradigms. However, architecture selection in modern enterprise environments continues to rely heavily on ad-hoc engineering intuition, with limited systematic frameworks to guide decision-making across heterogeneous database systems.

This paper introduces a unified cross-paradigm evaluation and selection framework for database architecture design in AI-ready data platforms. The framework is based on nine architectural dimensions and incorporates a structured multi-stage selection process involving workload characterization, constraint filtering, and compatibility scoring to enable systematic comparison and decision-making.

To ground the framework, we conduct a structured comparative analysis across thirteen major database paradigms spanning transactional, analytical, and AI-oriented systems. This analysis reveals three recurring patterns in database evolution: decoupling of storage and compute, workload-driven specialization, and convergence toward integrated AI-ready platforms.

The proposed framework is demonstrated through a representative enterprise case study in financial fraud detection, illustrating how hybrid, polyglot architectures emerge as optimal solutions for multidimensional workload requirements.

The cross-paradigm analysis culminates in an AI-ready reference architecture that integrates lakehouse storage, feature processing, and semantic retrieval layers as the unified substrate for modern analytics, machine learning, and Retrieval-Augmented Generation applications.

## I. INTRODUCTION

Modern enterprise data platforms operate in a complex design space shaped by polyglot data management and the integration of analytics, machine learning, and AI-driven workloads. Systems must simultaneously support transactional integrity, large-scale analytical processing, and semantic retrieval over high-dimensional data representations. Despite this complexity, database architecture selection in practice remains largely driven by ad-hoc engineering intuition, with limited systematic frameworks to guide decision-making across heterogeneous database paradigms.

This paper introduces a unified cross-paradigm evaluation and selection framework for database architecture design in AI-ready data platforms. The framework evaluates database systems across nine architectural dimensions, including data model, consistency, scaling strategy, storage layout, query model, workload fit, latency profile, cost model, and operational complexity, and incorporates a structured multi-stage decision process based on workload characterization, constraint filtering, and compatibility scoring.

To ground this framework, the paper analyzes the architectural evolution of database systems as a sequence of adaptations to changing workload characteristics, hardware capabilities, and system scale constraints. Rather than viewing database evolution as a linear progression, this study demonstrates that modern enterprise architectures are the result of accumulated specialization and convergence across multiple paradigms. This perspective aligns with foundational critiques of monolithic database architectures, which highlighted the limitations of one-size-fits-all systems [1].

Unlike prior surveys that examine database paradigms in isolation, this work provides a unified architectural evaluation across paradigms and synthesizes recurring convergence patterns toward AI-ready data platforms. The framework is further demonstrated through a representative enterprise case study, illustrating how hybrid, polyglot architectures emerge as optimal solutions for multidimensional workload requirements.

## II. METHODOLOGY: LITERATURE SELECTION AND ANALYSIS FRAMEWORK

This study employs a structured literature selection and analysis methodology consistent with established survey practices in database systems research. The methodology is designed to ensure systematic and representative coverage across foundational, widely adopted, and emerging paradigms, enabling both cross-paradigm architectural comparison and the development of a structured database architecture selection framework. The objective is to capture database systems, architectural paradigms, and selection methodologies that have materially influenced the evolution of enterprise data platforms, rather than to provide exhaustive enumeration of all published systems.

**Data Sources:**

The literature survey was conducted using major academic repositories, including IEEE Xplore, ACM Digital Library, SpringerLink, and Elsevier ScienceDirect, along with selected preprint sources for emerging AI-related topics.

**Search Keywords:**

Search queries addressed two complementary objectives. For architectural paradigm coverage, queries included combinations of terms such as "database systems," "relational databases," "NoSQL," "NewSQL," "distributed databases," "columnar databases," "lakehouse," "vector databases," and "AI data platforms." For database selection and evaluation framework literature, additional queries included "database selection framework," "database architecture selection," "database model selection," "database decision guidance," "polyglot persistence," "workload-driven architecture selection," and "hybrid database architecture."

**Time Range:**

The primary focus was on sources published between 2000 and 2025 that illustrate the transition from classical relational systems to modern distributed and AI-oriented data platforms. Foundational earlier works were included where necessary to establish architectural context. For selection framework and decision guidance literature, no strict lower time bound was applied, as methodologically relevant work spans a broader period.

**Selection Criteria:**

Two complementary criteria sets were applied. For paradigm literature, included studies focus on database architecture, storage models, or system design, with emphasis on scalability, consistency, and workload specialization; application-specific literature was excluded. For evaluation and selection framework literature, included studies propose structured methodologies for comparing, evaluating, or selecting database architectures across heterogeneous paradigms, covering polyglot persistence, decision guidance frameworks, and multi-paradigm architectural evaluation. Along with peer-reviewed articles, industrial reports and surveys were selected to reflect current trends in cloud-native and AI-oriented platforms.

**Final Corpus:**

Fifty-one representative works were selected: approximately 38 addressing paradigm-specific architecture, storage models, and system design, and 13 addressing database evaluation and selection methodologies, including polyglot persistence frameworks and multi-paradigm architectural evaluation. This corpus spans both peer-reviewed literature and industrial reports, ensuring coverage of the architectural characterization and decision-making dimensions addressed in this paper.

**Analysis Approach:**

The literature analysis served two complementary purposes. First, each paradigm was characterized across nine architectural dimensions, data model, consistency, scalability, storage layout, query model, workload fit, latency profile, cost model, and operational complexity, derived from recurring system design characteristics identified across the surveyed literature. Second, prior work on database evaluation and selection methodologies was examined to identify common constraint structures, workload characterization approaches, and compatibility assessment patterns that informed the proposed three-stage selection framework. These two analytical streams together underpin the nine-dimension evaluation model, the workload-relative selection framework, and the representative enterprise case study demonstrating its application.

## III. RELATED WORK

Prior research on database systems is broadly categorized into three streams: foundational system architecture, paradigm-specific surveys, and emerging cloud-native and AI-oriented data platform studies.

Foundational work on database architecture established the internal structure of relational database management systems, including storage engines, query optimizers, transaction managers, and recovery subsystems, providing design principles that underpin most subsequent systems [2]. These works defined the baseline architectural abstractions upon which later paradigms evolved.

A second body of literature focuses on paradigm-specific analysis. Large-scale data system surveys classify databases based on data models, consistency guarantees, and distribution strategies [3]. Surveys of NoSQL examined key-value, document, and wide-column stores designed for horizontal scalability and schema flexibility [4]. Research on main-memory databases has explored memory-optimized storage layouts, compiled query execution, and hardware-aware concurrency control [5]. Time-series systems for high-velocity telemetry [6] and graph databases for relationship-centric workloads [7] are part of additional surveys. These studies provide deep insight into individual paradigms but analyze systems in isolation without a unified cross-paradigm evaluation framework.

A third stream examines modern cloud-native and integrated data architectures. Studies on cloud databases have explored disaggregated storage and compute, elastic scaling, and serverless query processing [8], [9], while lakehouse research investigates how open table formats provide transactional guarantees over object storage [10]. Zhou et al.

[46] survey the intersection of databases and AI, focusing on how AI techniques can improve database internals such as query optimization and indexing rather than on how architectures should be selected and composed to support AI-driven workloads. While these works highlight important architectural convergence trends, each remains focused on specific platforms or technologies rather than providing a comprehensive cross-paradigm synthesis.

The most structurally relevant prior work on systematic database evaluation and selection consistently demonstrates paradigm-limited scope. Gessert et al. [48] provide a decision guidance framework for NoSQL systems, which evaluates key-value, document, columnar, and graph stores across functional and non-functional requirements. Davoudian et al. [49] survey NoSQL stores comprehensively across data models, consistency mechanisms, and architectural trade-offs. More recently, Roy-Hubara et al. [50] propose a structured database model selection method for polyglot persistence applications covering relational and NoSQL paradigms. Harby and Zulkernine [51] survey and experimentally compare data lake, warehouse, and lakehouse architectures. Each of these works shares a common limitation of addressing a specific paradigm class without providing a unified framework for cross-paradigm evaluation and selection.

Across all streams surveyed, to the best of our knowledge, no prior work provides a unified architectural perspective spanning the full evolution of database paradigms alongside a systematic framework for comparing and selecting architectures across heterogeneous systems. Importantly, such a framework need not position all paradigms as universal alternatives; rather, it must support workload-relative comparison in which constraint filtering eliminates architecturally unsuited paradigms before evaluation, ensuring that only relevant candidates are scored for a given use case. No prior work achieves this for systems that include AI-driven workloads integrating transactional systems, analytical platforms, and embedding-based retrieval.

## IV. MOTIVATION: DATABASE EVOLUTION IN THE ERA OF AI

Database systems have evolved from isolated transactional repositories into integrated platforms that support both analytical processing and AI-driven workloads. The explosion of data spanning structured, unstructured and semi-structured formats, is the primary reason for this transformation, rendering traditional siloed architectures insufficient.

Modern databases have evolved into AI-ready platforms that integrate vector search and embedding-based retrieval to support Retrieval-Augmented Generation (RAG), grounding large language models in fresh, proprietary context. Unified data lakehouse architectures that decouple compute from storage are required for enterprises.

This increases the complexity of database architecture design and necessitates careful consideration of consistency, scalability, latency, and workload specialization. It becomes necessary, therefore, to have methodologies through which database architectures can be determined methodically instead of depending upon intuition.

## V. CONTRIBUTIONS

This paper makes the following contributions:

- Introduces a unified cross-paradigm evaluation and decision framework for database architecture selection, based on a three-stage process of workload profiling, constraint filtering, and compatibility scoring (Section IX-B).
- Develops a nine-dimension architectural evaluation model covering data model, consistency, scalability, storage layout, query model, workload fit, latency, cost, and operational complexity. This enables systematic comparison across heterogeneous database systems.
- Provides a consolidated cross-paradigm analysis (Table II, Section VIII) which includes architectural trade-offs across major database paradigms within a single comparative reference.
- Identifies three recurring convergence patterns in modern data platforms, those are, progressive decoupling of storage and compute, workload-driven specialization, and synthesis of multiple paradigms within unified AI-ready architectures.
- Reinterprets database evolution as a sequence of architectural adaptations driven by workload requirements, system scale, and hardware constraints, supporting a unified view of modern polyglot data platforms.

## VI. EVALUATION FRAMEWORK

### *A. EVALUATION DIMENSIONS (TAXONOMY)*

This framework provides comparability among different paradigms through a single, multi-dimensional view. To enable consistent cross-paradigm comparison, this paper evaluates each database architecture across nine dimensions that together capture the full range of architectural characteristics governing system behavior under different workload conditions. The first, data model, concerns the structural representation of stored information - whether relational tables, document collections, graph structures, time-series sequences, or high-dimensional vector embeddings. The second, consistency and transaction model, defines the guarantees a system provides for concurrent reads and writes, spanning the spectrum from full ACID serializability through tunable BASE models to eventual consistency. Scaling strategy addresses the mechanism by which a system handles growth, distinguishing between vertical scale-up, horizontal scale-out, shared-nothing distribution, and distributed consensus approaches. Storage layout captures the physical organization of data on disk or in memory - row-store, column-store, LSM-tree, ANN index, or RAM-resident, a choice that has profound implications for which workloads a system can serve efficiently. The query model dimension characterizes the interface and semantics through which data is accessed, from SQL declarative and graph traversal languages to API-based access patterns and ANN similarity interfaces.

The remaining four dimensions address the operational and economic context of deployment. Workload fit identifies the operational profile for which a paradigm is architecturally optimized, OLTP, OLAP, HTAP (Hybrid Transaction/Analytical Processing), streaming, or AI and RAG workloads. Latency profile characterizes the responsiveness a system delivers for its primary operations, a dimension that varies by orders of magnitude across paradigms. Cost model captures the economics of compute and storage provisioning, distinguishing between tightly coupled on-premises infrastructure, cloud-native decoupled architectures, and per-byte consumption models. Operational complexity, the final dimension, reflects the engineering overhead required to deploy, tune, govern, and maintain the system at production scale.

The proposed nine architectural dimensions are derived by synthesizing recurring system design characteristics observed across the surveyed literature, rather than being arbitrarily defined, enabling consistent cross-paradigm comparison. These nine dimensions are applied consistently throughout the paradigm analyses in Section VIII, forming the basis for architectural characterization of each system.
The use of multiple dimensions reflects the inherent trade-offs across consistency, scalability, cost, and query expressiveness in database architectures. There are always trade-offs between consistency, scalability, costs, and query languages in any architecture advancements. This is precisely why polyglot persistence has become an architectural norm in enterprise data platforms rather than an exception.
It is important to note that these nine dimensions are not intended to position all paradigms as universal alternatives to one another. Database paradigms represent distinct architectural responses to different workload constraints rather than interchangeable options. The dimensions provide a common vocabulary for workload-relative comparison, with the constraint filtering stage of the selection framework ensuring that only paradigms architecturally suited to a given workload proceed to scoring.

### *B. REPRESENTATIVE PERFORMANCE CHARACTERISTICS*

Table I presents quantitative performance ranges across paradigms synthesized from distributed systems benchmarks, vendor documentation, and peer-reviewed database system evaluations. These ranges are directional rather than controlled benchmark results; actual performance varies significantly by implementation and deployment configuration. The quantitative characteristics presented here complement the qualitative architectural evaluation consolidated in Table II, presented at the conclusion of Section VIII

**TABLE I**
**REPRESENTATIVE PERFORMANCE CHARACTERISTICS ACROSS DATABASE PARADIGMS**

| Paradigm | Typical Write Throughput | Query Latency | Horizontal Scalability | Consistency Model | Storage Efficiency | Typical Node Scale |
|---|---|---|---|---|---|---|
| Relational (RDBMS) | $10^3$–$10^4$ ops/sec | Low | Limited | Strong (ACID) | High | Single → tens |
| NoSQL | $10^5$–$10^6$ ops/sec | Medium | Massive | Eventual / Tunable | Moderate | Hundreds–thousands |
| NewSQL | $10^4$–$10^5$ ops/sec | Medium (commit latency under distribution) | High | Strong Distributed ACID | High | Hundreds |
| Columnar | $10^3$–$10^4$ rows/sec | Very Low (analytical scans) | High | Strong | Very High | Hundreds |
| In-Memory | $10^5$–$10^6$ ops/sec | Ultra-Low (nanosecond range) | Moderate | Strong | Low (RAM cost) | Tens |
| Time-Series | $10^6$ writes/sec | Low (time-range queries) | Massive | Tunable | Very High | Thousands |
| Graph | $10^3$–$10^4$ traversals/sec | Very Low (relationship queries) | Moderate | Strong | Moderate | Tens–hundreds |
| Lakehouse | $10^3$–$10^4$ rows/sec (batch ingest); up to $10^6$ rows/sec (streaming ingest via distributed messaging layers) | Low–Very Low (analytical queries) | Massive | ACID via table formats | Very High | Hundreds–thousands |
| Vector DB | $10^4$–$10^5$ vectors/sec | Low (ANN search) | Massive | Eventual | High | Thousands |

Table I summarizes representative performance ranges across database paradigms, highlighting trade-offs between throughput, latency, and consistency under varying workload conditions. Values are synthesized from distributed systems benchmarks, vendor documentation, and peer-reviewed studies [17]–[20], [22], [24], [25] including representative systems across each paradigm. Performance ranges vary significantly across implementations and deployment configurations; values shown are directional rather than controlled benchmark results. These values should be interpreted as order-of-magnitude indicators rather than precise benchmarks.

## VII. BACKGROUND: ARCHITECTURAL DRIVERS AND HISTORICAL PROGRESSION

### *A. ARCHITECTURAL DRIVERS OF DATABASE EVOLUTION*

The historical progression of database systems is not merely a sequence of technological improvements but rather the result of recurring architectural pressures created by changing workloads, hardware capabilities, and system scale. Over many years of academic research and commercial database development, certain common drivers have emerged that motivate the introduction of new paradigms and obsolescence of older architectures.

#### DATA VOLUME AND DISTRIBUTION

With increasing data scale, ranging from megabytes to petabytes and even greater, vertical scaling was no longer adequate. To cope with this problem of scalability, distributed storage systems came into being, and they are also called shared-nothing architectures.

**WORKLOAD DIVERSITY**

Relational systems optimized for OLTP proved fundamentally inefficient for analytical workloads requiring high-throughput scanning and aggregation, driving the development of column-oriented OLAP systems [36]. This led to the rise of HTAP systems that attempted to achieve both in one single architecture.

**CONSISTENCY AND AVAILABILITY TRADE-OFFS**

There are unavoidable trade-offs among consistency, availability, and partition tolerance in any distributed system, as described by the CAP theorem [12], [13]. NoSQL systems prioritized availability over consistency to meet web-scale demands; NewSQL systems subsequently restored strong transactional guarantees through distributed consensus, accepting commit latency as the cost.

**HARDWARE EVOLUTION**

Advances in hardware have continued to transform database architecture. Advances in large-scale memory enabled the emergence of in-memory databases, and the availability of vectorized CPU instructions enabled columnar batch processing. Cloud computing has also enabled the creation of elastic compute and object storage services, which provide independent scalability for storage and compute resources.

**DATA VARIETY AND SCHEMA FLEXIBILITY**

Modern applications generate large volumes of semi-structured and unstructured data such as JSON documents, logs, and sensor streams. Rigid relational schemas proved difficult to adapt to these formats, leading to the emergence of schema-flexible document and wide-column data models.

**AI AND SEMANTIC WORKLOADS**

Machine learning and Generative AI workloads introduced new requirements for semantic retrieval over high-dimensional embeddings. Vector databases address this need by enabling approximate nearest neighbor search, extending traditional data platforms with embedding-based retrieval capabilities.

### B. TIMELINE OF DATABASE ARCHITECTURE EVOLUTION

Figure 1 summarizes the major phases of database architecture development from the 1960s to the present. Hierarchical and network models emerged in the 1960s for structured enterprise record management. The relational model was introduced in the 1970s which introduced SQL and ACID compliance as enduring standards. The 1990s brought column-oriented analytical systems and object-oriented databases, the mid-2000s produced distributed NoSQL systems prioritizing horizontal scale and availability, and the 2010s saw NewSQL restore distributed ACID guarantees alongside the emergence of lakehouse architectures unifying scalable storage with warehouse-grade reliability. Most recently, vector databases have extended enterprise platforms to support embedding-based semantic retrieval for AI applications.

This progression demonstrates that database evolution is characterized not by replacement of older paradigms but by gradual expansion of the ecosystem. Modern enterprise data platforms frequently incorporate multiple paradigms simultaneously, reflecting the increasing diversity of application workloads and architectural requirements, a pattern whose implications are analyzed in Section VIII.

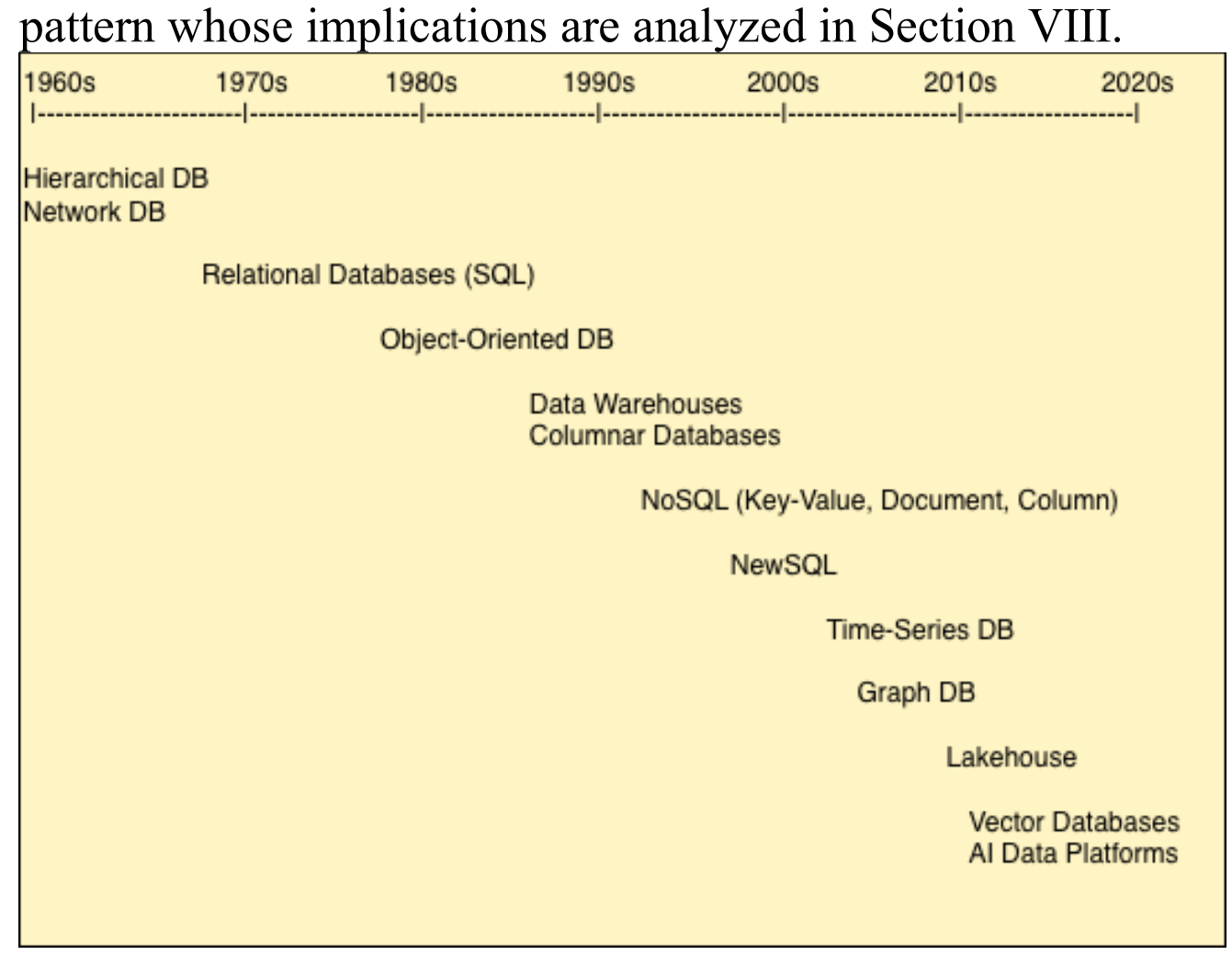

Fig. 1. Timeline of major database architecture paradigms illustrating the evolution from hierarchical data management systems to modern AI-ready data platforms integrating distributed analytics, lakehouse architectures, and vector retrieval systems.

## VIII. EVOLUTION OF DATABASE TECHNOLOGIES

This section examines the evolution of database technologies through a four-part analytical lens: Architectural Motivation, Architectural Innovations, Technical Constraints, and Workload Fit. Core paradigms are analyzed in detail, while earlier or specialized paradigms are treated in condensed form. Table II consolidates all paradigms within the unified evaluation framework.

### A. HIERARCHICAL AND EARLY PRE-RELATIONAL SYSTEMS

The first standardized database systems emerged in the 1960s with IBM's IMS and the CODASYL network model, followed by Object-Oriented databases (OODBs) in the 1980s. Although architecturally different, these two models also have a common limitation, which directly led to the development of the relational model: the strong association between data structure and physical storage.

Object-oriented databases were proposed as a solution for the impedance mismatch between object-oriented programming languages and relational storage by supporting native object persistence. However, the absence of a standardized declarative query language and limited interoperability restricted adoption to niche domains such as CAD/CAM and scientific computing.

### B. THE RELATIONAL REVOLUTION: DECLARATIVE LOGIC AND SCHEMA RIGOR

**Architectural Motivation:** The increasing complexity of enterprise applications in the 1970s was hindered by the inflexibility in navigability in hierarchical systems [14]. Applications needed to be able to query data in an unconstrained manner, without knowledge of access paths, and schema changes needed to be accommodated with architectural flexibility that was beyond pointer-based storage systems. E.F. Codd's relational model of 1970 [15] addressed both issues by using a mathematical model based on set theory, providing logical data independence as the guiding architectural principle.

**Architectural Innovations:** The relational model provided data independence, decoupling logical data view from physical storage, thus allowing for flexible schema change without requiring changes to applications. SQL provided a declarative query language, in which the optimizer, not the application, would determine query execution, thus moving the complexity boundary from application to database. ACID compliance was established as the benchmark for transactional integrity [15], [16] and normalization eliminated the update anomalies endemic to hierarchical systems by enforcing referential integrity through foreign key constraints.

**Technical Constraints:** Relational databases were architecturally vertical scale-up rather than scale-out. Preserving ACID integrity in a distributed environment introduced prohibitive complexity, particularly for multi-table joins at web scale where the cost of coordination across nodes grew non-linearly with data volume. The row-store layout, while optimal for OLTP point reads and writes, proved highly inefficient for analytical workloads requiring full-table scans across large datasets, as every column in every row must be read even when only a small subset of attributes is needed. These two constraints, horizontal scalability and analytical query performance are what drove every subsequent paradigm.

**Workload Fit and Adoption Context:** The relational model established itself as the sole source of truth for enterprise operations, driving complex ERP and CRM applications that required high-quality data for enterprise reporting and compliance. The success and dominance of relational databases from the 1970s through the 2000s reflects genuine architectural fitness for structured transactional workloads at moderate scale. It remains the primary paradigm for OLTP applications today, though it is increasingly complemented by analytical, distributed, and AI-oriented layers in modern enterprise architectures.

### C. THE NOSQL ERA: DISTRIBUTED SYSTEMS AND THE CAP THEOREM

**Architectural Motivation:** The mid-2000s data explosion, driven by web-scale platforms such as Google (Bigtable [17]) and Amazon (Dynamo [18]), exposed the vertical scaling limits of relational systems. Social media, e-commerce, and IoT platforms required the ability to distribute data across thousands of commodity nodes with continuous availability across geographies. At the same time, the rise of modern applications led to the requirement of handling semi-structured and unstructured data, JSON documents, user activity logs, and sensor data, which are difficult and expensive to handle in relational databases.

**Architectural Innovations:** NoSQL systems introduced diverse data models, document (JSON-based), key-value, wide-column, and graph alongside BASE consistency (Basically Available, Soft state, Eventual consistency) as a formally recognized alternative to ACID [33]. This shift replaced schema-on-write with schema-on-read, removing the requirement of expensive data migrations in agile development methodologies. The CAP theorem [12], [13] provided the theoretical framework that justified this trade-off: under network partitions, distributed systems must trade-off between consistency and availability under partition conditions, and NoSQL systems made this choice explicitly in favor of availability as illustrated in Fig.2.

**Technical Constraints:** Achieving strong consistency in a globally distributed NoSQL deployment incurs significant latency penalties, making these systems unsuitable for workloads requiring strict transactional guarantees such as financial records or inventory management [11]. The lack of a standardized query language for different NoSQL databases resulted in vendor lock-in and a higher learning curve for developers compared to SQL [35]. High-cardinality relational queries and complex joins on multiple entities result in poor performance compared to RDBMS systems and are hence less applicable for structured analytical applications.

**Workload Fit and Adoption Context:** NoSQL systems enabled globally distributed, highly available applications at web scale while prioritizing developer agility through schema flexibility. These trade-offs directly influenced subsequent paradigms, including NewSQL's restoration of transactional guarantees and the schema-flexible foundations of modern lakehouse architectures.

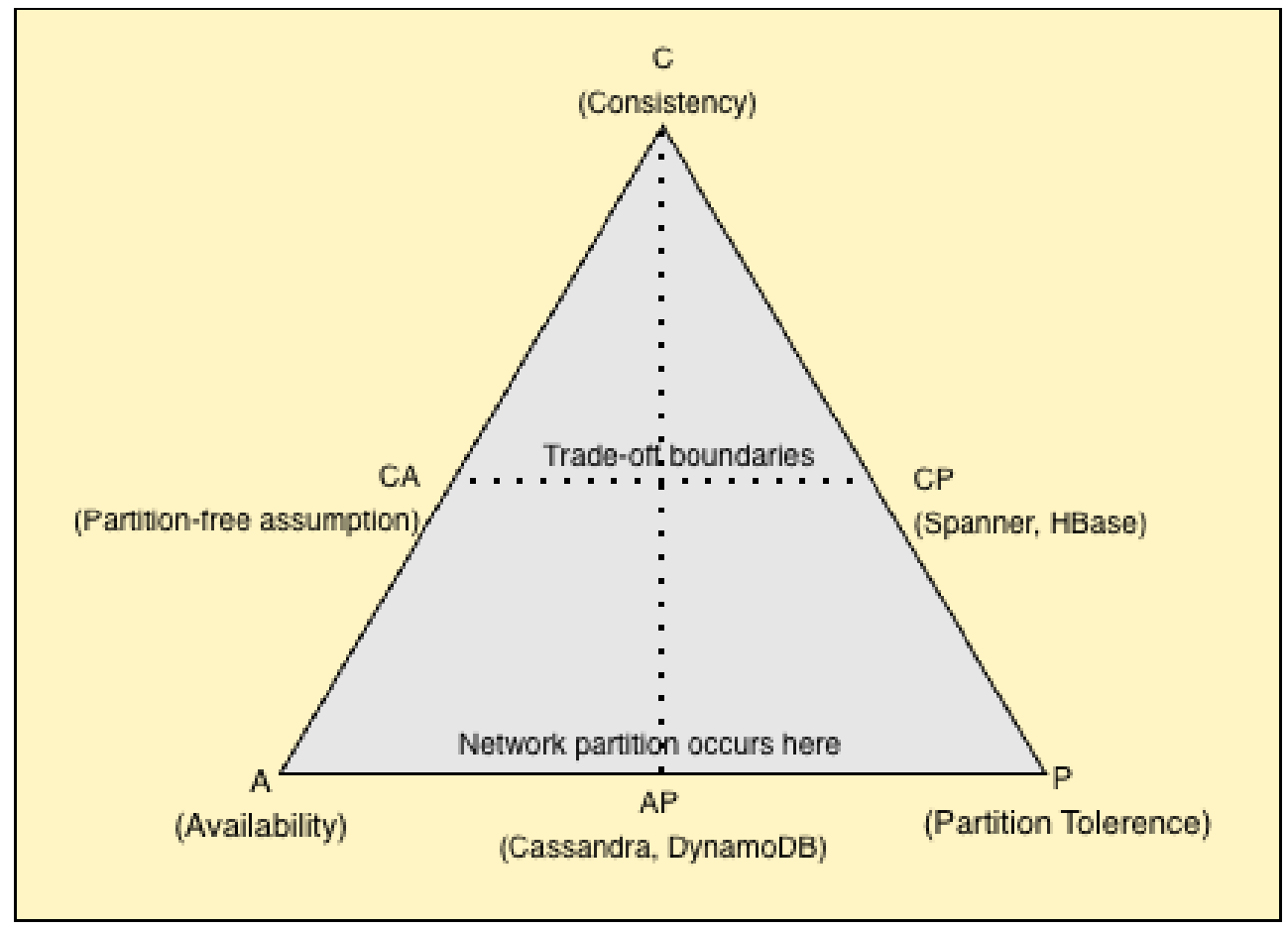

Fig. 2. Conceptual illustration of CAP trade-offs in distributed databases: under network partitions, designs typically prioritize consistency (CP) or availability (AP); CA is achievable only when partitions are absent

### D. COLUMNAR DATABASES: THE SHIFT TO ANALYTICAL THROUGHPUT

**Architectural Motivation:** The growth of business intelligence workloads in the late 1990s revealed a fundamental inefficiency in row-store systems. Analytical queries against wide tables typically access a small fraction of available columns, yet a row-store must scan all columns in every row touched by the query. For a 100-column table, a query needing 4 columns reads approximately 96% unnecessary data as shown in Fig.3. As datasets grew to terabytes, this I/O inefficiency made interactive analytical querying economically and technically impractical on relational infrastructure.

**Architectural Innovations:** Columnar systems inverted the storage layout, organizing data by attribute rather than record so that queries read only the columns they need, reducing I/O volume proportionally to the ratio of accessed columns to total columns [19], [20]. This storage inversion enabled two further innovations: vectorized query execution, in which CPU SIMD instructions process batches of column values simultaneously rather than row-by-row, and extreme compression ratios achieved by exploiting the homogeneity of column data through Run-Length Encoding (RLE), Delta Encoding, and dictionary compression [21]. The combination of reduced I/O, vectorized execution as illustrated in Fig.4, and compression transformed multi-terabyte analytical query latency from hours to seconds.

**Technical Constraints:** The same storage layout that optimizes analytical reads makes point writes expensive, inserting a single row requires updating multiple distinct column files, creating write amplification that makes columnar systems unsuitable for high-frequency transactional workloads. Wide SELECT queries are penalized because the engine must reconstruct full rows by stitching disparate column segments together at query time. These constraints are architectural, not merely design limitations, and represent fundamental trade-offs that cannot be changed without negating the analytical query performance benefits of the storage layout.

**Workload Fit and Adoption Context:** Columnar storage has become the core of modern cloud data warehouses, supporting complex trend analysis, prediction, and interactive visualization on multi-terabyte data sets without pre-computed cubes. The workload fit for columnar storage is limited, where OLAP performance is maximized at the expense of OLTP support, but within the OLAP workload, they are architecturally near-optimal. Cloud data warehouses have overcome the cost model of the original columnar storage paradigm, where compute and storage were tightly coupled, making the paradigm financially viable.

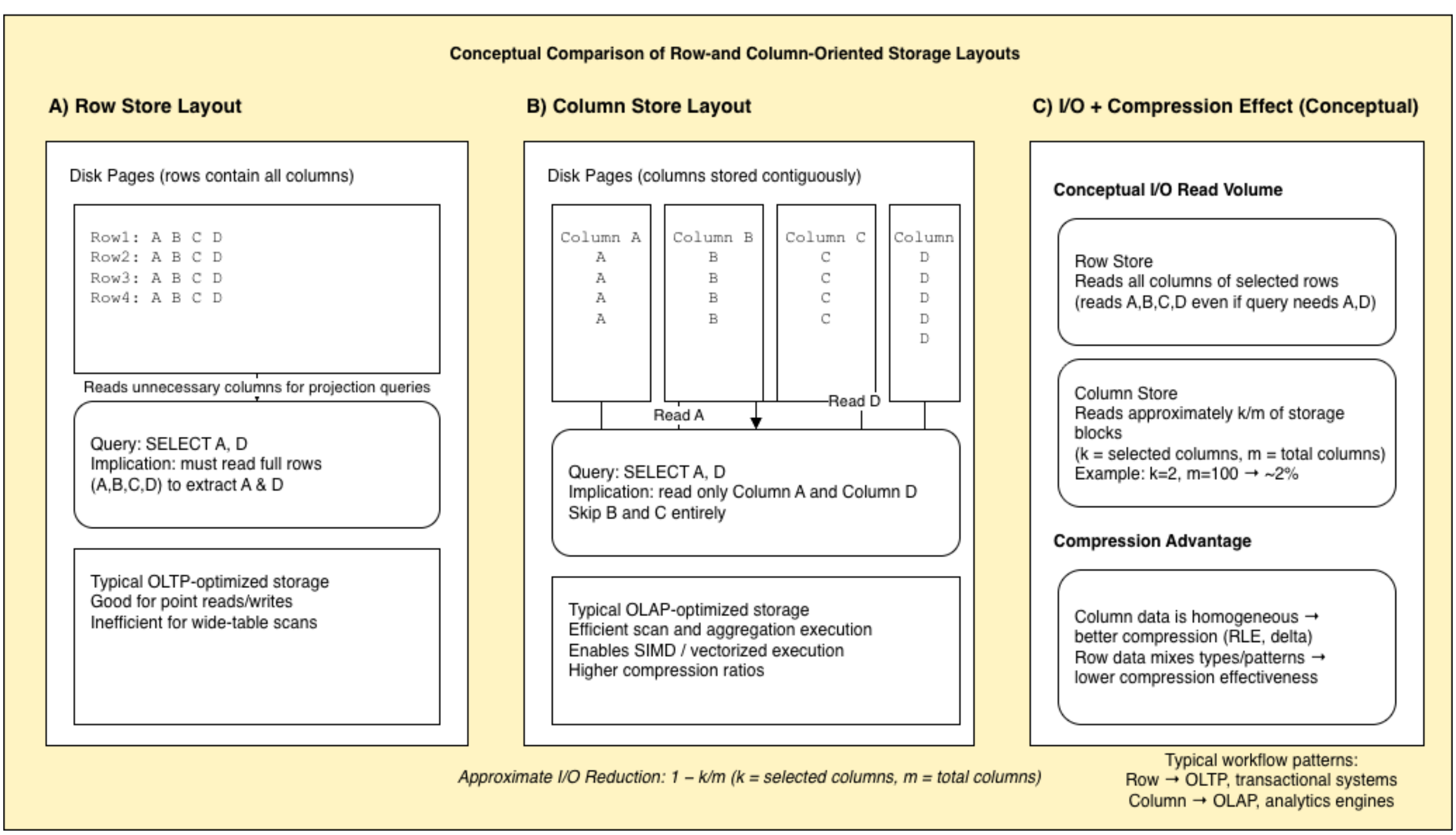

Fig. 3. Row-store vs column-store access pattern for projection-heavy analytical queries; column stores reduce I/O by scanning only selected attributes and benefit from higher compressibility of homogeneous columns

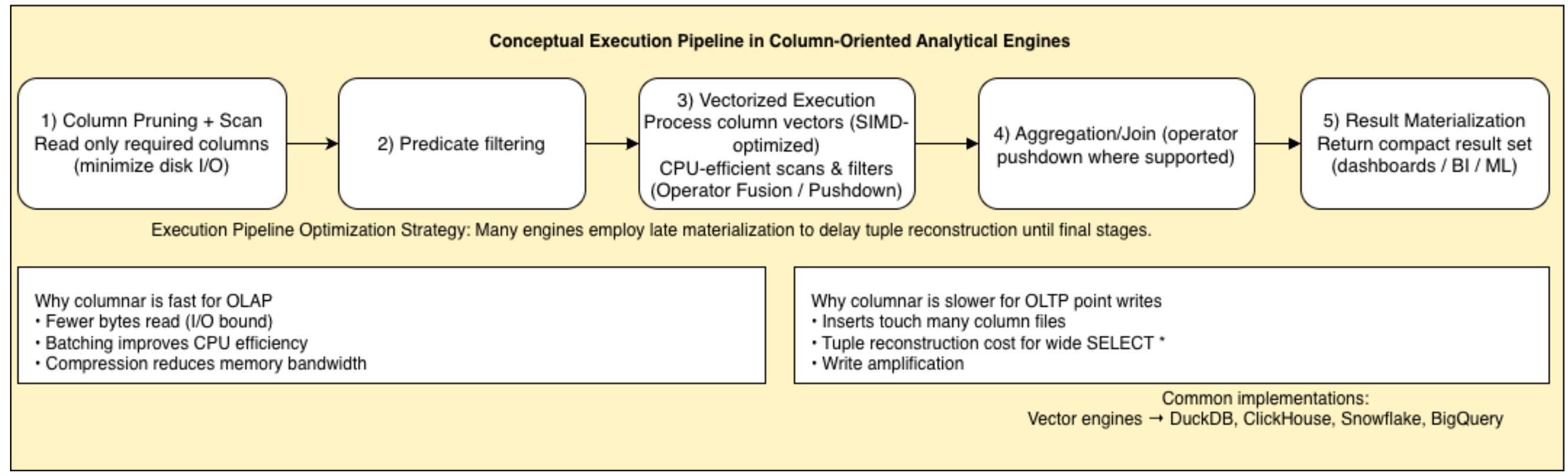

Fig. 4. Conceptual columnar query execution pipeline highlighting column pruning, predicate pushdown, vectorized execution, and late materialization, key mechanisms behind high OLAP throughput.

### *E. NEWSQL: DISTRIBUTED ACID AT SCALE*

**Architectural Motivation:** By the early 2010s, a class of applications had emerged that required properties neither RDBMS nor NoSQL could provide simultaneously: the strict transactional guarantees and SQL expressiveness of relational systems, and the horizontal scalability and geographic distribution of NoSQL. Global financial platforms, large-scale SaaS applications, and multinational transaction processing systems could not accept the consistency trade-offs of NoSQL but had outgrown the vertical scaling ceiling of traditional RDBMS. NewSQL emerged to resolve this issue directly.

**Architectural Innovations:** NewSQL systems including Google Spanner [22], CockroachDB, and TiDB reimagined the relational engine for distributed environments using distributed consensus protocols - Paxos or Raft, to ensure consistency across nodes without a single central coordinator. The use of globally synchronized clocks (TrueTime in Spanner) or logical clocks has facilitated the ordering of transactions at the global level, thereby allowing the logical database to span multiple nodes and continents with strict serializability [23]. Automatic data partitioning, replication, and rebalancing between clusters have removed the need for manual data sharding, which made horizontal scaling of the traditional RDBMS operationally infeasible.

**Technical Constraints:** The requirement for distributed ACID properties implies waiting for quorum acknowledgement of all nodes for every transaction before committing, which leads to commit latency. The sensitivity of NewSQL to network conditions, due to the aforementioned latency, makes them more network condition dependent than either single-node RDBMS or eventually consistent NoSQL databases. The infrastructure cost of distributed consensus is also much higher for NewSQL compared to the infrastructure cost of single-node RDBMS, making the cost of NewSQL databases justifiable only for environments where scale and consistency are hard requirements.

**Workload Fit and Adoption Context:** NewSQL is architecturally appropriate for the specific intersection of globally distributed scale and non-negotiable transactional integrity, financial systems, compliance-regulated platforms, multinational SaaS applications, rather than as a general-purpose replacement for RDBMS or NoSQL. Its demonstration that distributed ACID is technically achievable at global scale directly influenced the transactional guarantee mechanisms subsequently implemented by lakehouse open table formats such as Delta Lake.

### *F. IN-MEMORY DATABASES: ELIMINATING THE I/O BOTTLENECK*

**Architectural Motivation**: Traditional disk-based databases are, by nature, constrained by I/O latencies and are not appropriate for ultra-low-latency applications such as high-frequency trading or fraud detection. Even the most advanced storage devices, such as NVMe, are not able to provide the necessary microsecond-level latencies that are required. Therefore, as memory became not only affordable but also spacious, in-memory databases were born. In-memory databases are focused on the system's RAM as the storage device, not as a secondary cache.

**Architectural Innovations:** IMDBs utilize specialized lock-free data structures and indexes specifically engineered for RAM-resident data, eliminating the buffer pool management overhead that disk-based systems require. Systems such as SAP HANA [47] extended this model with columnar in-memory formats that enable simultaneous OLTP and OLAP processing within a single engine, the HTAP pattern. By serving both transactional and analytical workloads against the same in-memory data store, these systems remove the need for the dedicated ETL pipelines that traditional architectures require to move data from operational systems into separate analytical platforms, significantly reducing latency between data capture and analytical availability.

**Technical Constraints:** Despite their performance advantages, IMDBs face significant technical hurdles. The volatility of DRAM necessitates complex logging and synchronous replication to ensure data durability, an engineering overhead largely avoided by disk-persistent systems. Furthermore, the high per-byte cost of RAM limits the economic viability of IMDBs to datasets that can fit within a defined memory budget. Finally, "warm-up" latency, the time required to reload data into RAM after a system restart, poses a substantial risk to high-availability service level agreements (SLAs).

**Workload Fit and Adoption Context:** IMDBs are reserved for latency-critical applications where sub-millisecond response requirements cannot be met by disk-based alternatives regardless of optimization. The HTAP capability pioneered by columnar in-memory systems influenced subsequent hybrid architectures and established real-time analytical processing on operational data as an architectural goal that lakehouse and HTAP platforms later pursued at larger scale and lower cost, trading nanosecond latency for petabyte capacity.

### G. TIME-SERIES DATABASES: ENGINEERING FOR HIGH-CARDINALITY STREAMS

**Architectural Motivation**: The rise of microservices observability and Industrial IoT platforms in the 2010s created ingestion and retention patterns that general-purpose databases were structurally unsuited to handle – producing millions of timestamped data points from high-cardinality sources in continuous append-only streams that exposed core inefficiencies in relational and NoSQL databases. Time series databases like InfluxDB, Prometheus, and TimescaleDB were invented directly in response, with every architectural layer optimized for the most common operation, high-frequency sequential writes with time-bound range queries.

**Architectural Innovations:** TSDBs adopted LSM-Tree (Log-Structured Merge-Tree) [32] storage layouts as their foundational design choice, converting random writes into sequential append operations that minimize write amplification under sustained high-frequency ingestion. This storage model is complemented by time-partitioned compression schemes that exploit the temporal locality and numerical regularity of timestamped data to achieve compression ratios significantly beyond what general-purpose engines can deliver. Down sampling capabilities and adjustable data retention rules help solve the issue of storage scalability by incrementally aggregating the historical data at increasingly coarse-grained time intervals without any input from users.

**Technical Constraints:** Performance degrades rapidly when tag cardinality, the number of unique metadata dimensions associated with time-series, grows into the millions, creating index bloat that undermines the ingestion efficiency the architecture is designed to deliver. Relational join support is absent, and complex updates to historical records are poorly supported, making TSDBs unsuitable as standalone systems for workloads requiring dimensional analysis against slowly-changing reference data.

**Workload Fit and Adoption Context:** TSDBs deliver unmatched ingestion throughput for observability, predictive maintenance, and IoT workloads where the primary operation is continuous high-velocity append with range-based retrieval. Their deliberate sacrifices, eventual rather than strong consistency, absent relational expressiveness are acceptable in monitoring contexts where minor data loss is tolerable and queries are time-bounded. For workloads requiring both high-velocity ingestion and relational dimensional analysis, hybrid architectural patterns combining TSDBs with relational or lakehouse layers are required, as discussed in Section IX-C.

### H. VECTOR DATABASES: THE SEMANTIC RETRIEVAL LAYER FOR AI

**Architectural Motivation:** The emergence of large language models and embedding-based machine learning introduced a retrieval requirement that no preceding paradigm was designed to address: finding semantically similar items across billions of high-dimensional numerical vectors in millisecond timeframes. Semantic similarity across embedding spaces requires a fundamentally different index structure and query model. The growing deployment of LLMs in enterprise environments made this capability not a research curiosity but a production infrastructure requirement.

**Architectural Innovations:** Vector databases store data as high-dimensional numerical embeddings generated by ML models and retrieve results using Approximate Nearest Neighbor (ANN) algorithms [45], most notably HNSW (Hierarchical Navigable Small World graphs [24]), that navigate embedding spaces through a layered graph structure, achieving millisecond-range similarity search at billion-vector scale without exhaustive distance computation [34]. The primary operational application is Retrieval-Augmented Generation: vector databases serve as the external semantic memory through which LLMs retrieve contextually relevant enterprise content at inference time, grounding model responses in current proprietary data without retraining.

**Technical Constraints:** The cost structure of vector databases scales non-linearly with embedding dimensionality, storage requirements and query computation both grow with the number of dimensions, while the curse of dimensionality simultaneously degrades the discriminative power of distance metrics beyond certain thresholds. Vector freshness presents an ongoing operational challenge: as source documents are updated, corresponding embeddings must be regenerated and reindexed to prevent semantic drift between the system of record and the retrieval layer. Consistency is typically eventual, and query expressiveness outside ANN operations is limited, making vector databases unsuitable as primary systems of record and confining their role to the semantic retrieval layer within broader polyglot architectures.

**Workload Fit and Adoption Context:** Vector databases are not replacements for existing paradigms but required complements, a semantic retrieval layer that extends relational, analytical, and lakehouse platforms to support AI workloads. The hybrid architectural patterns required to combine vector retrieval with transactional consistency are examined in Section IX-C.

### I. LAKEHOUSE ARCHITECTURE AND OPEN TABLE FORMATS

**Architectural Motivation:** By the mid-2010s, enterprises working with data lakes and data warehouses found that there was a structural redundancy issue. Data warehouses had schema enforcement and good query performance, but they were expensive and not flexible enough with regards to handling semi-structured data. Data lakes, which provided excellent object storage capabilities for semi- and fully-structured data, lacked strong data governance and query performance capabilities. This created significant engineering overhead and introduced consistency risks between maintaining both systems.

**Architectural Innovations:** Lakehouse architectures combined both of these paradigms through the use of open table formats such as Delta Lake [25], [26], Apache Iceberg, and Apache Hudi, which add transactional logs, snapshot isolation, schema evolution, and metadata governance directly on top of low-cost distributed object storage. The defining architectural innovation is compute-storage decoupling: multiple analytical engines, SQL processors, ML frameworks, and streaming pipelines [31] operate against a single governed storage layer without data duplication or movement. This principle has direct architectural lineage to the MapReduce distributed batch processing model [27], which first demonstrated that analytical computation could scale horizontally across commodity infrastructure independently of storage, a pattern that Apache Spark [28] subsequently extended with in-memory DAG execution and that lakehouse architectures now generalize to multi-engine, multi-workload environments. Time-travel queries, the ability to query data as it existed at any prior point in time, are enabled by the transaction log maintained by open table formats, providing both auditability and recovery capabilities.

**Technical Constraints:** Lakehouse platforms require mature data engineering practices and orchestration tooling that single-paradigm systems do not. Managing open table format compaction, partition optimization, and metadata scaling at enterprise data volumes demands dedicated engineering investment. Batch analytical query latency, while significantly improved over raw data lake approaches, remains higher than purpose-built data warehouse systems for certain query patterns.

**Workload Fit and Adoption Context:** Lakehouse platforms achieve the broadest workload fit of any paradigm surveyed, supporting analytical processing, ML feature pipelines, streaming ingestion, and AI workloads against a single governed storage layer. Their cloud-native cost model, with independent compute and storage scaling and pay-per-use economics, has made them the dominant design pattern for modern enterprise data platforms. The integration of vector retrieval layers into lakehouse infrastructure represents the current leading edge of this convergence, positioning lakehouses as the unified substrate for the AI-ready reference architecture described in Section VIII-K.

### J. ADDITIONAL PARADIGMS: GRAPH, MULTI-MODEL, SPATIAL, AND BLOCKCHAIN

**Graph Databases.** Graph databases optimize relationship-centric workloads by storing edges as first-class structures and enabling efficient multi-hop traversal queries [7]. However, dense graph structures create partitioning challenges that limit horizontal scalability. Graph databases are architecturally essential for relationship-centric workloads, real-time fraud detection, master data management, knowledge graph construction where no other paradigm approaches their traversal latency profile.

**Multi-Model Databases.** Multi-model databases like ArangoDB and Azure Cosmos DB combine document, graph, and key-value models in a single database system [39], thus reducing database sprawl associated with implementing dedicated database systems for specific data models. A single storage system supports multiple query APIs, thus eliminating ETL processing for moving data between dedicated systems. This approach also reduces associated costs for licensing, infrastructure, and human resources. The architectural trade-off is that supporting multiple data models in a single system means forgoing optimizations that are specific to a given workload, thus establishing performance limits for any given dimension. Consistency and scaling modes are flexible rather than fixed. This approach is most suitable for applications where diversity and ease of management outweigh performance in any single dimension.

**Spatial Databases.** Spatial databases such as PostGIS and Esri add an index structure to relational or NoSQL database systems, which allows for fast geographic queries. The index structures used in spatial databases, such as the R-Tree and Quadtree, divide space into boxes, enabling the database to quickly exclude irrelevant geographic areas and only consider relevant coordinates [40]. This enables efficient proximity, containment, and intersection queries finding delivery vehicles within a defined radius, calculating route intersections, environmental boundary analysis that general-purpose B-Tree indexes cannot support efficiently. Consistency model and scaling strategy inherit from the host engine; the architectural contribution is a specialized storage layout and query model for location-intelligence workloads where geographic operations are primary rather than peripheral.

**Blockchain Databases.** The unique position of blockchain databases is that their consistency model is able to provide cryptographic immutability through the use of decentralized consensus algorithms such as Proof of Work, Proof of Stake, or PBFT in a manner that no other paradigm is capable of [41], [42]. Data is organized into cryptographically linked blocks with Merkle root verification, creating a tamper-evident append-only ledger that enables trustless multi-party data sharing without a central intermediary. This guarantee comes at the cost of the lowest write throughput and highest latency of any architecture surveyed, with minimal query expressiveness and a cost model with no analog in conventional database economics, every participating node stores and validates the full chain. Workload fit is trust-oriented rather than performance-oriented, justified specifically where the cost of a trusted central intermediary exceeds the overhead of distributed consensus, supply chain provenance, decentralized finance, and regulated multi-party audit trails.

### K. REFERENCE ARCHITECTURE FOR AI-READY DATA PLATFORMS

An AI-ready data platform integrates six architectural layers into a cohesive operational system. The Operational Data Layer consists of transactional systems - relational databases, event streams, and application stores, that serve as the authoritative system of record. The Ingestion and Streaming Layer captures batch and real-time data through distributed messaging and change data capture pipelines, decoupling producers from downstream consumers. The Unified Storage Layer, implemented as a lakehouse with open table formats, provides a single governed substrate supporting both structured and unstructured data with transactional guarantees and metadata governance. The Processing and Feature Layer applies distributed compute engines for analytics, feature engineering, and ML transformations against this shared storage layer. The Semantic Retrieval Layer provides vector indexing and ANN search, enabling semantic access to enterprise knowledge for AI applications. The Serving and Application Layer exposes processed data and model outputs through APIs, dashboards, AI services, and agent-based systems. As illustrated in Figure 5, the RAG workflow connects these layers concretely: documents are preprocessed, chunked, embedded, and indexed offline, while online inference embeds the user query, retrieves semantically relevant context via ANN search, and assembles an augmented prompt for LLM inference, with a feedback loop continuously improving retrieval relevance. This architecture represents a fundamental shift from passive data storage toward active computational substrates for intelligent decision-making.

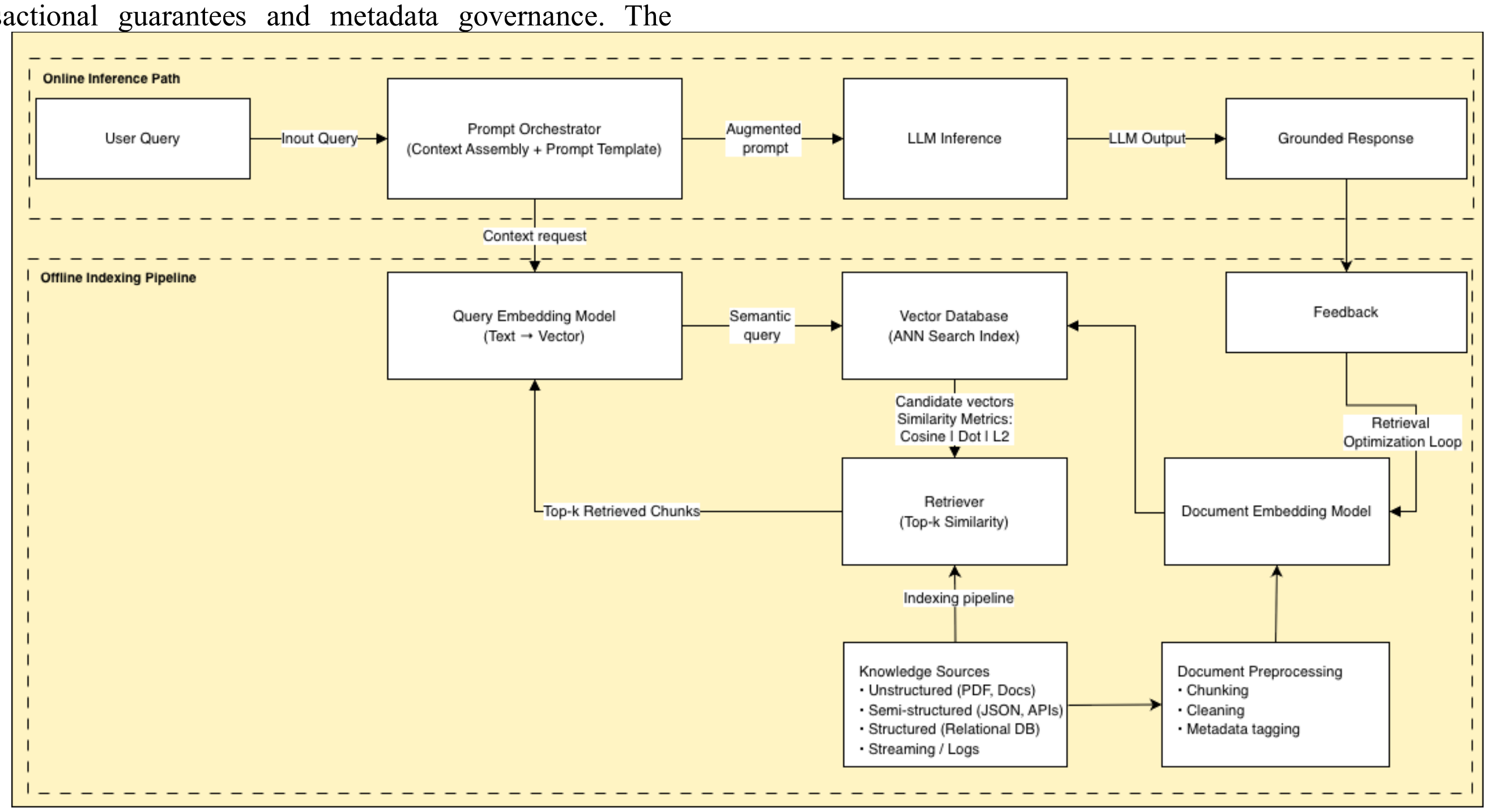

Fig. 5. Reference RAG architecture separating offline indexing (chunking, embedding, vector index build) from online inference (query embedding, ANN retrieval, context assembly, LLM inference) with feedback-driven retrieval optimization.

The reference architecture presented in this section integrates the paradigms surveyed throughout Section VIII into a unified operational model. Table II consolidates the nine-dimension architectural evaluation across all thirteen paradigms, providing the comparative foundation for the convergence analysis and selection framework that follows in Section IX.

TABLE II
CROSS-PARADIGM ARCHITECTURAL EVALUATION ACROSS NINE DIMENSIONS

| Paradigm | Data Model | Consistency Model | Scaling Strategy | Storage Layout | Query Model | Workload Fit | Latency Profile | Cost Model | Operational Complexity | Primary Trade-off |
|---|---|---|---|---|---|---|---|---|---|---|
| Hierarchical | Tree / navigational | Strong | Vertical only | Row, pointer-linked | Navigational | Narrow known-path OLTP | Low (known paths); high (schema change) | High TCO, tightly coupled | Low rigid, brittle to change | Rigid schema, no data independence |
| Relational | Relational tables | Strong (ACID) | Vertical, limited horizontal | Row-store | SQL declarative | OLTP, moderate analytical | Low (OLTP); high (full-table scans) | Coupled compute-storage | Medium | Join complexity at scale; analytical inefficiency |
| NoSQL | Document / KV / Wide-column | Tunable / BASE | Horizontal, massive | Document / KV / Wide-column | API / JSON / custom | Web-scale, semi-structured | Low–Medium write; variable read | Low commodity nodes | Medium | Consistency trade-offs; no standardized query language |
| Columnar | Column-oriented relational | Strong | Horizontal | Column-store | SQL (OLAP) | Analytical (OLAP) | Very low for scans; high for point writes | Medium historically coupled; cloud-native now decoupled | Medium | Slow point writes; OLTP unfit |
| NewSQL | Relational | Strong distributed ACID | Horizontal | Row-store | SQL | Distributed transactional | Medium commit latency under distribution | High consensus infrastructure cost | High clock management, network sensitivity | Commit latency; high operational cost |
| In-Memory | Relational / hybrid | Strong | Moderate, RAM-bounded | RAM-resident | SQL / API | OLTP + HTAP | Ultra-low (nanosecond range) | Very high DRAM cost per byte | High checkpointing, warm-up | Volatile; cost prohibitive at large scale |
| Time-Series | Timestamp-indexed | Tunable | Horizontal, massive | LSM-tree / delta-encoded | Time-range specialized | Streaming metrics, IoT, observability | Ultra-low for ingestion; low for range queries | Low extreme compression efficiency | Medium | High-cardinality degradation; no relational joins |
| Graph | Property graph | Strong | Limited horizontal | Index-free adjacency | Traversal (Cypher) | Relationship-centric analytics | Very low for path traversal | Medium scales poorly at high node counts | High graph sharding complexity | Partition complexity limits horizontal scale |
| Lakehouse | Multi-format structured + unstructured | Strong (via table formats) | Horizontal, massive | Columnar + object storage | SQL + ML APIs | Analytics, ML, streaming, AI/RAG | Low–Very low (analytical); near-real-time (streaming) | High efficiency decoupled, pay-per-use | High orchestration and data engineering maturity required | Operational complexity; batch query latency |
| Multi-Model | Document / graph / KV hybrid | Configurable | Horizontal | Hybrid | Multi-API | Mixed workloads | Medium across supported types | High consolidated TCO efficiency | Medium | Performance ceiling on any individual workload type |
| Spatial | Geometric / relational | Inherits host engine | Inherits host engine | R-Tree / Quadtree | Geometric SQL | Location intelligence | Low for proximity and containment queries | Inherits host; spatial index maintenance overhead | Medium | Coordinate system complexity; inherits host engine limitations |
| Blockchain | Append-only blocks | Cryptographic immutability | Decentralized, very limited | Replicated full chain | Minimal / smart contracts | Trust-critical, multi-party audit | Very high latency per transaction | Very high full chain replicated on every node | Very high enterprise integration friction | Throughput and latency severely constrained by consensus |
| Vector | High-dimensional embeddings | Eventual | Horizontal, massive | ANN index (HNSW) | Similarity / ANN search | Semantic AI retrieval, RAG | Low (millisecond ANN at billion-vector scale) | High index compute and dimensional scaling cost | Medium | Dimensional cost scaling; eventual consistency only |

Table II presents a consolidated evaluation of thirteen database paradigms across the nine architectural dimensions defined in Section VI-A, plus a Primary Trade-off summary column. The table is positioned at the conclusion of Section VIII so that each rating is grounded in the paradigm analyses presented in Sections VIII-A through VIII-K. Ratings reflect paradigm-level architectural tendencies rather than properties of any specific product implementation; individual systems may extend or partially mitigate the characteristics shown. For the Cost Model column, "High efficiency" denotes favorable economics relative to capability delivered; "High cost" denotes unfavorable or expensive economics. Paradigms rated "Inherits host engine" in the Spatial row reflect the architectural dependency of spatial extensions on an underlying platform whose characteristics vary by deployment.

## IX. ARCHITECTURAL SYNTHESIS: CONVERGENCE PATTERNS AND SELECTION FRAMEWORK

### *A. CONVERGENCE PATTERNS IN MODERN DATABASE ARCHITECTURES*

The paradigm-by-paradigm analysis in Section VIII reveals that database evolution has followed a pattern of accumulation and convergence rather than simple replacement, where newer paradigms address specific limitations of earlier systems while coexisting within broader enterprise architectures. Examining this evolution in aggregate reveals three recurring architectural patterns that characterize how database systems are currently converging toward unified AI-ready platforms.

**Decoupling as a Primary Architectural Driver**. A consistent pattern across modern database architectures is the progressive separation of tightly coupled components to enable independent scaling. The transition from hierarchical to relational systems decoupled logical data representation from physical storage. Lakehouse architectures extended this by decoupling compute engines from storage layers, enabling organizations to run SQL analytics, ML pipelines, and streaming workloads against a single governed data layer without duplication. The distributed batch processing model introduced by MapReduce [27] established the foundational precedent for this compute-storage separation, demonstrating at petabyte scale that computation and data need not be co-located to achieve fault-tolerant analytical throughput. Cloud-native database platforms have taken this further, separating transaction processing from query execution [8]. This decoupling pattern reflects a systematic response to the cost model constraints that emerge when compute and storage must scale together under workload heterogeneity.

**Specialization as a Complementary Force.** In parallel with convergence, many new paradigms achieve performance breakthroughs precisely by narrowing their workload scope. Columnar databases sacrifice write throughput for analytical scan efficiency. Time-series databases sacrifice relational expressiveness for ingestion velocity. Vector databases sacrifice consistency guarantees for semantic retrieval performance. Rather than representing architectural regressions, these specializations reflect deliberate trade-offs that make each paradigm exceptionally effective within its target workload. The consequence for enterprise architecture is that no single system can serve all workloads optimally, which drives the adoption of polyglot persistence as an architectural norm rather than an exception [29], [30].

**Convergence as the Emerging Synthesis.** The most recent architectural trajectory represented by lakehouse platforms, HTAP systems, and AI-ready data platforms reflects an attempt to synthesize the strengths of multiple paradigms within a unified infrastructure [46]. Modern platforms such as lakehouse architectures with open table formats combine ACID transactional guarantees, columnar analytical performance, streaming ingestion, and ML pipeline integration within a single governed storage layer [26]. The integration of vector retrieval layers into these platforms further extends this convergence to support Retrieval-Augmented Generation and semantic search alongside structured analytics [29]. These developments appear to point to the future direction of database architecture as being one where there is integration of data and AI platforms, where paradigm boundaries become less distinct, rather than any particular approach dominating the future direction of database architecture.

### *B. DATABASE ARCHITECTURE SELECTION FRAMEWORK*

Today, most enterprise systems do not rely on a single database paradigm; architects have to evaluate several architectural options based on workload characteristics, performance requirements, system constraints, and operational considerations. To support this decision process, this paper proposes a conceptual three-stage architectural selection framework that systematically evaluates database paradigms against workload characteristics. The evaluation uses the nine architectural dimensions defined in Section VI-A: data model, consistency guarantees, scaling strategy, storage layout, query model, workload fit, latency profile, cost model, and operational complexity. As illustrated in Fig. 6, the framework consists of three stages:

**Workload Profiling** – identifying workload requirements and priorities

**Constraint Filtering** – eliminating architectures that violate essential system constraints

**Weighted Compatibility Scoring** – ranking candidate architectures based on compatibility with workload requirements

These stages progressively narrow the candidate architectures and identify the database paradigms most suitable for the target workload.

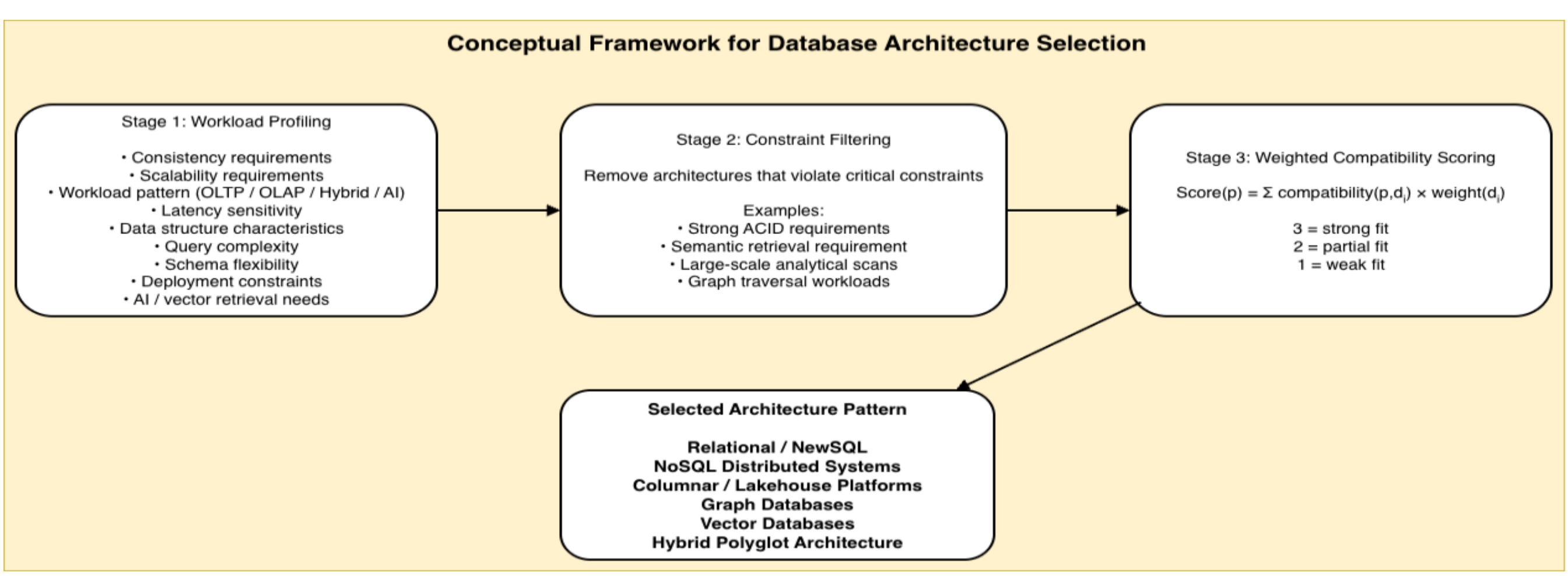

Fig. 6. Conceptual framework for database architecture selection based on workload profiling, architectural constraint filtering, and weighted compatibility scoring.

**Stage 1**: **Workload Profiling**. In the first stage, the architect profiles the target workload across a set of workload requirement dimensions that capture the operational and functional requirements of the system.

Typical workload dimensions include:

- consistency requirement level
- scalability requirement level

- workload pattern (OLTP, OLAP, hybrid, streaming, or AI pipelines)
- latency sensitivity
- data structure characteristics (tabular, document, graph, time-series, embeddings)
- query complexity and analytical depth
- schema flexibility requirements
- deployment and operational constraints
- AI or embedding-based retrieval requirements

For each dimension $d_i$, the architect assigns a priority weight $w_i$ representing its relative importance for the target system. Higher weights indicate dimensions that have greater influence on architecture selection.

Formally, the workload profile can be represented as:

$$W = \{(d_i, w_i, r_i)\}$$

Where,

- $d_i$ represents the workload dimension
- $w_i$ represents the importance weight
- $r_i$ represents the minimum requirement threshold

This stage produces a structured representation of workload priorities that guides the architectural evaluation process.

**Stage 2: Constraint Filtering.** In the second stage, candidate database paradigms are evaluated against critical architectural constraints derived from the workload profile. This stage applies hard constraints that eliminate paradigms which cannot satisfy non-negotiable system requirements. Database paradigms that fundamentally conflict with essential workload requirements are removed from the candidate set. This step prevents architectures that violate core system constraints from being considered in later stages of evaluation. The filtering operates at the level of database paradigms rather than specific implementations, and therefore reflects general architectural tendencies rather than strict rules

**TABLE III**
**REPRESENTATIVE ARCHITECTURAL CONSTRAINTS AND CANDIDATE PARADIGM REDUCTION**

| Workload Constraint | Primary Candidate Architectures | Architectures Generally Disfavored |
|---|---|---|
| Strong ACID guarantees required (financial or compliance workloads) | Relational DBMS, NewSQL | NoSQL systems emphasizing eventual consistency, vector databases as primary system of record |
| Native semantic or embedding-based retrieval required | Vector databases, analytical platforms with vector indexing extensions | Classic relational or NoSQL systems without vector indexing |
| High schema variability or schema-on-read analytics | NoSQL document stores, lakehouse platforms | Strongly typed transactional relational systems |
| Massive distributed scale with high write throughput | NoSQL distributed systems, NewSQL distributed databases | Single-node relational architectures |
| Large-scale analytical aggregation workloads | Columnar databases, lakehouse platforms | Row-store OLTP systems |
| Highly interconnected relationship analysis | Graph databases | Relational systems relying on heavy recursive joins |

Note: These constraints represent paradigm-level architectural tendencies rather than absolute limitations. Specific database implementations may extend or partially mitigate these constraints.

**Stage 3 - Weighted Compatibility Scoring.** After constraint filtering reduces the candidate architectures, the remaining database paradigms are evaluated using a weighted compatibility scoring model.

For each paradigm $p$, a suitability score is calculated as:

$$Score(p) = \sum_{i=1}^{n} compatibility(p, d_i) \times w_i$$

Where,

- $compatibility(p, d_i)$ represents how well database paradigm $p$supports workload dimension $d_i$
- $w_i$represents the importance weight assigned to that dimension in Stage 1.

Compatibility values are interpreted qualitatively as:

| Score | Interpretation |
|---|---|
| 3 | Strong architectural alignment |
| 2 | Partial or conditional fit |
| 1 | Weak alignment or significant trade-offs |

Compatibility scores are derived from the architectural analysis presented earlier in this paper, where each database paradigm is evaluated across the dimensions introduced in Section VI-A. The database paradigm with the highest aggregate score represents the best architectural fit for the workload. It is common that several paradigms may achieve high scores, which often leads to hybrid architectures combining specialized systems.

The framework is demonstrated through a detailed enterprise case study in Section IX-D, applying all three stages to a financial fraud detection platform with multidimensional workload requirements spanning transactional integrity, real-time decisioning, analytical processing, and semantic retrieval. The case study illustrates how the scoring model leads systematically to a polyglot architectural recommendation.

**Framework Limitations** The proposed framework is a conceptual architectural decision model operating at the paradigm level; it does not account for product-specific optimizations, deployment configurations, or hardware acceleration effects.

### *C. EDGE CASES AND HYBRID ARCHITECTURAL PATTERNS*

In real-world systems, there are instances when the requirements cross more than one paradigm at once, making hybrid/polyglot architecture unavoidable. The next examples present such cases when a singular paradigm cannot address the architectural problem, while a particular hybrid approach can.

**Edge Case 1 - Transactional Consistency with Semantic Retrieval (RAG over Enterprise Data).** The integration of Retrieval-Augmented Generation into enterprise workflows creates a workload pattern that no single paradigm was designed to support: an application must simultaneously maintain ACID-compliant transactional records (contracts, policies, financial instruments) and perform millisecond-range semantic similarity search over embedded representations of those same documents. Relational databases and NewSQL systems provide the required consistency guarantees but lack native Approximate Nearest Neighbor (ANN) indexing required for semantic retrieval. Vector databases offer the retrieval functionality, but they only offer eventual consistency models, which do not support the type of transactional guarantees that are necessary for a system of record. The recommended architectural solution is to use a hybrid solution, which is a relational or NewSQL database as a system of record, with a vector database as a semantically indexed read replica, using CDC pipelines to synchronize the two. The consistency boundary between the two stores must be explicitly governed, specifically, the propagation delay between a document update in the system of record and the corresponding embedding refresh in the vector index creates a window of semantic staleness that applications must account for. This operational constraint is distinct from the retrieval quality considerations addressed in the original RAG formulation [29] and reflects a systems engineering concern that arises specifically in enterprise deployments where the source data is transactionally managed and subject to frequent updates.

**Edge Case 2 - High-Cardinality Time-Series with Relational Joins.** Industrial IoT and observability platforms routinely require combining high-velocity sensor or telemetry streams with slowly-changing relational dimensions, asset registries, customer hierarchies, organizational metadata. Pure time-series databases handle ingestion and retention efficiently but lack native join support at relational complexity, while relational databases degrade under high-frequency append workloads. The recommended architectural response is a lakehouse ingestion pattern in which time-series data is continuously materialized into a lakehouse storage layer where it can be joined against dimensional tables using distributed SQL engines. This approach unifies the governance model within a single storage substrate at the cost of some ingestion latency, a trade-off acceptable in most observability and maintenance contexts.

**Edge Case 3 - Knowledge Graph with Vector Embedding Integration.** Graph databases and vector databases address complementary retrieval challenges, structured relationship traversal and semantic similarity search respectively, and an increasing class of enterprise AI applications requires both simultaneously. Knowledge graph applications in life sciences, financial compliance, and enterprise search must traverse typed relationships (entity → regulation → jurisdiction → exemption) while also identifying semantically similar entities across embedding spaces. Neither paradigm supports the other's query model natively. Emerging architectural patterns address this through hybrid graph-vector indexing, in which graph nodes carry embedding vectors that can be retrieved by ANN search and then expanded through graph traversal, or through unified query languages that combine graph traversal predicates with vector distance functions. This remains an active research area and represents one of the clearest directions for future architectural convergence at the intersection of knowledge representation and Generative AI.

These three edge cases reinforce the main observation of this survey: none of the database paradigms fully satisfy the multidimensional requirements of complex enterprise workloads. Each boundary scenario resolves through a hybrid architectural response, typically a combination of two paradigms with explicit management of the consistency, latency, and governance boundaries between them. The selection framework in Section IX-B should therefore be understood as yielding not a single paradigm recommendation but a primary paradigm and its required complements, which together constitute the full architectural response to a given workload profile.

### D. Case Study: Application of the Selection Framework

This section demonstrates the full three-stage application of the proposed framework to a financial fraud detection platform, a representative enterprise workload requiring simultaneous transactional integrity, real-time decisioning, large-scale analytical processing, and semantic similarity retrieval.

**Stage 1: Workload Profile**

Table IV presents the workload profile for this scenario. Consistency and latency receive the highest weights, reflecting the non-negotiable requirements for financial-grade transactional guarantees and sub-millisecond fraud signal generation. Workload fit and scaling strategy receive elevated weights due to the multi-paradigm nature of the workload. Remaining dimensions are weighted at standard importance.

TABLE IV
WORKLOAD PROFILE: FINANCIAL FRAUD DETECTION PLATFORM

| Dimension ($d_i$) | Weight ($w_i$) | Rationale |
|---|---|---|
| Consistency | 5 | Financial records require strict ACID compliance |
| Latency Profile | 5 | Real-time fraud detection requires sub-millisecond response |
| Workload Fit | 4 | Must support OLTP, analytics, and AI/semantic retrieval simultaneously |
| Scaling Strategy | 4 | High-volume, globally distributed transaction processing |
| Query Model | 3 | SQL for transactions; ANN similarity for fraud pattern detection |
| Data Model | 3 | Structured transaction records and high-dimensional embeddings |
| Storage Layout | 2 | Secondary to workload fit and consistency requirements |
| Cost Model | 2 | Enterprise budget with manageable infrastructure economics |
| Operational Complexity | 2 | Manageable complexity at enterprise production scale |
| Total | 30 | |

**Stage 2: Constraint Filtering**

Applying the hard constraints derived from the workload profile eliminates four paradigms from consideration. NoSQL systems (BASE/eventual consistency) are eliminated as primary systems for financial records. Time-series databases (tunable consistency, absent relational joins) are eliminated. Blockchain databases are eliminated due to impractical write throughput and transaction latency. Hierarchical systems are eliminated as architecturally obsolete for this workload class. Vector databases are conditionally retained as a semantic retrieval layer but are excluded from consideration as a primary system of record due to eventual consistency.

Three additional paradigms are eliminated on workload-specific grounds. Graph databases, while architecturally relevant for fraud-ring relationship traversal, are eliminated as primary candidates due to limited horizontal scalability that conflicts with the globally distributed, high-volume transaction processing requirement (weight 4); their complementary role is noted but does not warrant independent scoring in this workload profile. Multi-model databases are eliminated because their performance ceiling across any individual workload dimension makes them unsuitable where OLTP transactional integrity, large-scale analytics, and semantic retrieval each carry elevated weights simultaneously. Spatial databases are eliminated as the workload profile contains no geographic query requirement. The surviving paradigms that proceed to Stage 3 are: Relational, NewSQL, Columnar, In-Memory, Lakehouse, and Vector DB (conditional role).

**Stage 3: Weighted Compatibility Scoring**

Table V presents the compatibility scores for surviving paradigms across the nine architectural dimensions. Compatibility values are assigned as: 3 = strong architectural alignment, 2 = partial or conditional fit, 1 = weak alignment or significant trade-offs.

TABLE V
WEIGHTED COMPATIBILITY SCORING: FINANCIAL FRAUD DETECTION PLATFORM

| Paradigm | Data Model (3) | Consistency (5) | Scaling (4) | Storage Layout (2) | Query Model (3) | Workload Fit (4) | Latency (5) | Cost Model (2) | Op. Complexity (2) | Total Score |
|---|---|---|---|---|---|---|---|---|---|---|
| **Relational** | 3 | 3 | 2 | 3 | 3 | 2 | 3 | 2 | 2 | **78** |
| **NewSQL** | 3 | 3 | 3 | 2 | 3 | 2 | 2 | 1 | 1 | **71** |
| **In-Memory** | 2 | 3 | 2 | 3 | 2 | 3 | 3 | 1 | 1 | **72** |
| **Columnar** | 2 | 3 | 3 | 3 | 2 | 1 | 1 | 2 | 2 | **62** |
| **Lakehouse** | 3 | 3 | 3 | 3 | 3 | 3 | 2 | 3 | 1 | **81** |
| **Vector DB*** | 2 | 1* | 3 | 3 | 1 | 3 | 3 | 2 | 2 | **67*** |

*Vector DB consistency score reflects eventual consistency model. Retained as semantic retrieval layer only; excluded as primary system of record per Stage 2 constraint filtering.

Score = $\Sigma$ compatibility($p$, $d_i$) × $w_i$; maximum possible score = 90.

**Results and Architectural Recommendation**

The scoring reveals three distinct clusters. Lakehouse (81) and Relational (78) achieve the highest aggregate scores, reflecting strong alignment with the dominant workload requirements. In-Memory (72) and NewSQL (71) score closely, representing viable alternatives where ultra-low latency or global distribution respectively take priority. Vector DB (67), though constrained by eventual consistency, scores highest on semantic workload fit and retrieval latency, confirming its role as the required semantic layer. Columnar (62) is effectively subsumed within the Lakehouse's columnar storage layer and does not warrant independent deployment.

These results lead directly to a polyglot architecture composed of three complementary systems: a Relational or NewSQL database as the primary transactional system of record; a Lakehouse platform for large-scale analytical processing and machine learning pipelines; and a Vector database as a semantically indexed retrieval layer, synchronized via CDC pipelines with the transactional system. This outcome confirms the central observation of this framework: for multidimensional enterprise workloads, systematic scoring does not converge on a single paradigm but instead identifies the primary paradigm and its required architectural complements.

## X. INDUSTRY ADOPTION AND ARCHITECTURAL DIRECTION

Recent industry reports indicate that spending on cloud database platforms and enterprise Data and AI platforms is gaining momentum, reflecting the increasing convergence of analytics and AI application workloads in production environments [37], [38]. These trends point to an ongoing shift in database architecture towards mixed data environments, where transactional storage, analytics platforms, and AI-oriented retrieval layers (such as embedding-based retrieval and vector indexing) are integrated to enable real-time analytics and RAG-style applications [29], [43], [44]. Taken together, these adoption patterns reinforce the central architectural observation of this paper: modern database ecosystems are evolving toward integrated, workload-aware platforms rather than single-paradigm data infrastructures.

## XI. LIMITATIONS

This study focuses on architectural paradigms rather than product-level benchmarking. Performance characteristics vary significantly across implementations, deployment environments, workloads, and hardware configurations. Accordingly, the performance ranges presented in Table I should be interpreted as directional estimates synthesized from representative literature and publicly reported system characteristics, rather than results derived from controlled experimental evaluation. The purpose of this comparison is to provide architectural intuition across paradigms rather than definitive performance ranking.

The proposed framework operates at the level of architectural abstraction. It does not provide empirical benchmarking or quantitative validation across specific system implementations. It is intended as a decision-support model grounded in qualitative synthesis of recurring system characteristics observed across the literature. This positioning reflects the goal of enabling cross-paradigm comparison and architectural reasoning rather than system-level performance optimization.

In practice, database paradigms are not strictly mutually exclusive. Modern data platforms frequently integrate multiple paradigms, such as relational systems supporting semi-structured data or analytical platforms incorporating vector indexing. The classifications used in this paper should be interpreted as architectural abstractions not rigid system boundaries.

Finally, new architectural patterns will continue to be introduced due to rapid changes in database technologies. Future work may extend the proposed framework to incorporate empirical validation and adapt to these evolving paradigms.

## XII. CONCLUSION

This paper introduces a unified evaluation and selection framework for designing database architectures in AI-ready data platforms. Rather than viewing database evolution as a sequence of replacements, this study demonstrates how new paradigms emerge as adaptations to shifting workload requirements. The analysis identifies three dominant architectural patterns: architectural decoupling (for scalability), workload specialization (for paradigm-level performance), and architectural convergence (for integrating specialized systems).

As organizations adopt machine learning pipelines, vector retrieval, and real-time decisioning systems, the challenge shifts from selecting a single database to determining an appropriate combination of complementary systems. Database architecture design therefore becomes a multi-dimensional decision problem requiring systematic evaluation across heterogeneous solutions. The proposed framework addresses this challenge by facilitating structured comparison and informed selection of database paradigms.

Future work can extend this framework through empirical validation and the automation of a decision-support system. Overall, this work establishes a foundation for systematic database architecture design in evolving AI-driven data environments.